\documentclass[fleqn,twoside]{article}
\usepackage{espcrc2}

\usepackage{graphicx}
\usepackage[figuresright]{rotating}

\newcommand{\AmS}{{\protect\the\textfont2
  A\kern-.1667em\lower.5ex\hbox{M}\kern-.125emS}}
\def\lsim{\;\raise0.3ex\hbox{$<$\kern-0.75em\raise-1.1ex\hbox{$\sim$}}\;}
\def\gsim{\;\raise0.3ex\hbox{$>$\kern-0.75em\raise-1.1ex\hbox{$\sim$}}\;}
\def\bea{\begin{eqnarray}}   \def\eea{\end{eqnarray}}
\def\beq{\begin{equation}}   \def\eeq{\end{equation}}
\def\nn{\nonumber}

\def\nl{\newline}

\title{Phenomenological Aspects of the
Next-to-Minimal Supersymmetric Standard Model}

\author{Ulrich Ellwanger\address[MCSD]{Laboratoire de Physique
Th\'eorique, CNRS -- UMR 8627 \\ Universit\'e de Paris XI, F-91405
Orsay Cedex, France
\\ \hspace*{1cm} \\LPT Orsay 09-66}}
       
\begin{document}

\begin{abstract}
The motivations for the NMSSM are reviewed, and possible unconventional
signals for Higgs and sparticle production at the LHC are discussed. In
the presence of a light pseudoscalar, the SM-like Higgs scalar can decay
dominantly into a 4-tau final state. In the fully constrained NMSSM with
mSUGRA-like soft SUSY breaking terms, the correct dark matter relic
density is obtained for a singlino-like LSP which modifies considerably
all sparticle decay chains.
\vspace{1pc}
\end{abstract}

\maketitle

\section{The NMSSM}

The Next-to-Minimal Supersymmetric Standard Model (NMSSM) addresses the
so-called $\mu$-problem of the MSSM~\cite{Kim:1983dt}, whose origin we
describe below: 

Any supersymmetric extension of the Standard Model (SM) generalizes in a
unique way -- as dictated by supersymmetry -- the interactions involving
dimensionless gauge- and Yukawa couplings, whereas the electroweak
scale originates from the softly supersymmetry breaking mass terms and
trilinear interactions (of the order $M_\mathrm{SUSY}$).  The MSSM,
however, requires the introduction of a \emph{supersymmetric} (SUSY)
mass term for the Higgs multiplets, the so-called $\mu$-term: both
complex Higgs scalars $H_u$ and $H_d$ of the MSSM have to be components
of chiral superfields which contain, in addition, fermionic
$SU(2)$-doublets $\psi_u$ and $\psi_d$. Some of the $SU(2)$-components
of $\psi_u$ and $\psi_d$ are electrically charged. Together with with
the fermionic superpartners of the $W^\pm$~bosons, they constitute the
so-called chargino sector (two charged Dirac fermions) of the SUSY
extension of the SM. Due to the fruitless searches for a chargino at
LEP, the lighter chargino has to have a mass above $\sim 103$~GeV
\cite{LEPSUSYWG}. Analysing the chargino mass matrix, this lower limit
implies that a Dirac mass $\mu$ for $\psi_u$ and~$\psi_d$ -- for
arbitrary other parameters -- has to satisfy the constraint $|\mu| \gsim
100$~GeV. A Dirac mass term is not among the soft supersymmetry breaking
mass terms, hence $\mu$ has to be a \emph{supersymmetric} mass term for
the Higgs multiplets.

In addition, an analysis of the Higgs potential shows that a
non-vanishing soft SUSY breaking term $B\mu\,H_u H_d$ is a necessary
condition so that \emph{both} neutral components of $H_u$ and $H_d$ are
non-vanishing at the minimum. This, in turn, is required in order to
generate masses both for up-type quarks and down-type quarks by the
Higgs mechanism. The numerical value of $B\mu$ should be
roughly of the order of the electroweak scale~($M_Z^2$).

However, $|\mu|$ must not be too large: the Higgs potential must be
unstable at its origin $H_u = H_d = 0$ in order to generate the
electroweak symmetry breaking. Whereas negative soft SUSY breaking mass
terms for $H_u$ and $H_d$ of the order of the SUSY breaking scale
$M_\mathrm{SUSY}$ can generate such a desired instability, the
$\mu$-induced masses squared for $H_u$ and $H_d$ are always positive,
and must \emph{not} dominate the negative soft SUSY breaking mass terms.
Consequently the $\mu$~parameter must obey $|\mu| \lsim
M_\mathrm{SUSY}$. Hence, both ``natural'' values $\mu = 0$ and $\mu$
very large ($\sim M_\mathrm{GUT}$ or $\sim M_\mathrm{Planck}$) are ruled
out, and the need for an explanation of $\mu \approx M_\mathrm{SUSY}$ is
the $\mu$-problem.

Within the NMSSM, $\mu$ is generated in a way similar to the quark- and
lepton masses with the help of a vacuum expectation value (vev) of a
scalar field: one introduces a Yukawa coupling $\lambda$ of the
higgsinos $\psi_u$ and~$\psi_d$ to a scalar field $S$ (a gauge singlet,
since the $\mu$-parameter carries no gauge quantum numbers), and
arranges that the vev $\left< S\right>$ is of the order of
$M_\mathrm{SUSY}$. This is easy to obtain with the help of soft SUSY
breaking negative masses squared (or trilinear couplings) of the order
of $M_\mathrm{SUSY}$ for $S$; then, $M_\mathrm{SUSY}$ is the only scale
in the theory. In this sense, the NMSSM is the simplest supersymmetric
extension of the SM in which the weak scale is generated by the
supersymmetry breaking scale $M_\mathrm{SUSY}$ only.

It should be mentioned that additional attractive features of the MSSM
-- the unification of the running gauge couplings at $M_\mathrm{GUT}
\sim 10^16$~GeV and the possibility to explain the dark matter relic
density -- remain unchanged.

\subsection{The NMSSM Superpotential and Particle Content}

The superpotential $W_{NMSSM}$ of the NMSSM depends on the additional
gauge singlet superfield~$S$, and is obtained from the
superpotential $W_{MSSM}$ of the MSSM by the following substitutions:
\bea
W_{MSSM} &=& {\mu} H_u H_d\ + \dots\nn\\
\to 
W_{NMSSM}& =& {\lambda} S H_u H_d 
+\frac{1}{3}{\kappa}S^3\ + \dots
\eea
and similarly for the soft SUSY breaking terms:
\beq
{B \mu} H_u H_d\ + \dots \to
{\lambda A_\lambda} S H_u H_d 
+\frac{1}{3} {\kappa A_\kappa} S^3\
+ \dots
\eeq

The term $\sim \kappa$ in $W_{NMSSM}$ serves to stabilize the potential
for the scalar singlet field $S$; a possibly negative SUSY breaking
mass term $m_S^2 |S|^2$ (together with ${\frac{1}{3}\kappa A_\kappa}
S^3$+ h.c.) can easily generate a vev $\left< S\right> \sim
M_\mathrm{SUSY}$. Comparing $W_{NMSSM}$ and $W_{MSSM}$ one finds an
effective $\mu$-term with $\mu_\mathrm{eff} = \lambda \left< S\right>$.

Discarding the Goldstone bosons, the particle content of the
Higgs sector of the NMSSM consists in 3 CP-even neutral scalars $H_i$,
2~CP-odd neutral scalars $A_i$, 1 charged Higgs scalar $H^\pm$, and --
together with the neutral gauginos -- 5 neutralinos $\chi_i^0$. (The
singlet-like states $H_S$, $A_S$ and $\chi^0_S$ mix with the
$SU(2)$-doublets and the neutral gauginos; the decomposition of the
eigenstates depend on the parameters $\lambda$, $\kappa$, $A_\lambda$
and $A_\kappa$.)

It is important to note that the lightest CP-odd scalar $A_1$ can be
quite light, notably in the case of an approximate $R$-symmetry in the
Higgs sector ($A_\lambda$, $A_\kappa \to 0$) or Peccei-Quinn-symmetry
($\kappa \to 0$), in which case it plays the r\^ole of a (pseudo-)
Goldstone boson.

\section{Phenomenological Aspects of the NM\-SSM}

As a result of the new states and mixings in the Higgs and
neutralino sectors, the corresponding phenomenology of the NMSSM can
differ considerably from the MSSM. Already the LEP constraints from
Higgs boson searches have to be interpreted anew.

\subsection{Lessons/Hints from LEP}

The results of the four LEP experiments sear\-ching for a Higgs scalar
decaying into $H \to b\bar{b}$, $\tau^+\,\tau^-$ (assuming SM branching
fractions) have been combined by the LEP-Higgs Working
Group~\cite{Schael:2006cr} and are shown in~Fig.~1. There, $\xi$ denotes
the reduced coupling of a Higgs scalar to the $Z$ boson (compared to the
coupling of the SM Higgs scalar), $\xi \equiv g_{HZZ}/g_{HZZ}^{SM}$.
Shown are upper bounds on $\xi^2$ as function of a scalar Higgs mass
$m_H$.

\begin{figure}[h!]
\includegraphics[scale=0.37]{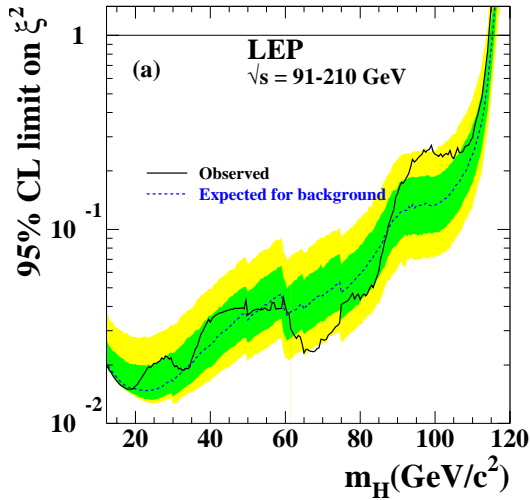}
\caption{ Upper bound on $\xi^2$ as function of a scalar Higgs mass
$m_H$, where $\xi$ denotes the coupling of the Higgs scalar to $Z$
bosons (normalized w.r.t. the SM Higgs boson);
from~\cite{Schael:2006cr}.}
\end{figure}
%\vspace*{-6mm}

One can note a light excess of events for $m_H \sim 95-100$~GeV (of
$\sim 2.3\, \sigma$ statistical significance), which is difficult to
explain in the SM. The NMSSM offers two possible explanations for this
excess of events: i) a Higgs scalar with a mass of $\sim
95-100$~GeV can have a reduced coupling to the $Z$ boson
($\xi \lsim 0.4-0.5$) due to its large
singlet component; or ii) a Higgs scalar with a mass of $\sim
95-100$~GeV can have a reduced branching ratio into $b\bar{b}$,
$\tau^+\,\tau^-$, since it decays dominantly into a pair of light CP-odd
scalars with a $BR(H\to A_1 A_1)\sim 80-90\%$. In the latter case, the
coupling of $H$ to $Z$ bosons can be SM-like.
In~\cite{Dermisek:2005gg,Dermisek:2006wr}, it has been argued that this
scenario allows to alleviate the ``little finetuning problem'' of
supersymmetric extensions of the SM (since $m_H \gsim 114$~GeV is not
required).

However, the LEP experiments have also sear\-ched for $H \to A_1 A_1 \to
4b$~\cite{Schael:2006cr}, and the constraints are very strong for
$m_H \sim 95-100$~GeV. On the other hand, if $M_{A_1}$ is below the
$b\bar{b}$ threshold of $\sim 10.5$~GeV, $A_1$ would decay dominantly
into $\tau^+\,\tau^-$. LEP constraints on $H \to A_1 A_1 \to
4\tau$ impose no bounds for $m_H \sim 95-100$~GeV~\cite{Schael:2006cr};
hence $M_{A_1}\lsim 10.5$~GeV is an attractive scenario.

\subsection{Lessons/Hints from radiative $\Upsilon$ decays}

For $M_{A_1} \lsim 10.5$~GeV, decays $\Upsilon(nS) \to A_1 + \gamma$ are
kinematically possible. The corresponding branching ratios depend on
$M_{A_1}$, and the coupling $g_{A_1 b \bar{b}}$ of $A_1$ to $b$-quarks.
It is useful to define a reduced coupling $X_d\equiv g_{A_1 b \bar{b}}/
g_{H b \bar{b}}^{SM}$, and for large $\tan\beta$ one can obtain $X_d >
1$ in the NMSSM.

In the range $M_{A_1}\lsim 9$~GeV where $A_1 \to\tau \tau$ (or $A_1
\to\mu \mu$) would be dominant, the CLEO collaboration has searched for
$\Upsilon(1S) \to A_1 + \gamma$ decays~\cite{:2008hs}. No signal has
been observed, which implies upper bounds on $X_d$ dependent on
$M_{A_1}$. 

The BABAR collaboration has recently observed an  $\eta_b$-like state
with a mass of $\simeq 9.39$~GeV in $\Upsilon(2S)$ and $\Upsilon(3S)$
radiative decays~\cite{Aubert:2008vj,:2009pz}. For large $g_{A_1 b
\bar{b}}$, the $\eta_b(nS)$ $b\bar{b}$ bound states ($n=1,2,3$) can mix
with $A_1$, since these states have the same quantum
numbers~\cite{Drees:1989du,Fullana:2007uq,Domingo:2008rr,Domingo:2009tb},
and the mixing elements/eigenvalues of the $\eta_b(nS)- A_1$ mass matrix
can be determined in terms of $X_d$ and 
$M_{A_1}$~\cite{Fullana:2007uq,Domingo:2008rr,Domingo:2009tb}.

The measured mass by BABAR of the $\eta_b$-like state must correspond to
one of these eigenvalues. On the one hand, this implies that $M_{A_1}$
cannot be equal to $9.39$~GeV, if the mixing (which is
proportional to $X_d$) is large: for a large mixing, an eigenvalue of a
$2\times 2$ matrix (of the $\eta_b(1s)-A_1$ system) cannot coincide with
one of its diagonal elements. This reasoning implies an upper bound on
$X_d$ for $M_{A_1}$ near 9.39~GeV. In Fig.~2 we show upper bounds on
$X_d$ in the NMSSM from CLEO (black), the muon anomalous magnetic moment
$a_\mu$ (blue), $B$-physics (green) and constraints due to the measured
$\eta_b(1S)$ mass by Babar as a red line (from~\cite{Domingo:2008rr}).

\begin{figure}[h!]
\vspace*{3mm}
\includegraphics[scale=0.37]{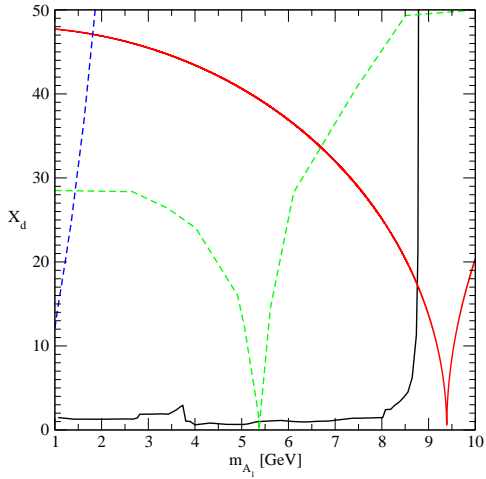} \caption{Upper bounds on
$X_d$ versus $M_{A_1}$ in the NMSSM. Indicated are constraints from
$B$-physics as a green dashed line, constraints from $a_\mu$ as a
blue dashed line, the bounds from CLEO on $BR\left(\Upsilon \to
\gamma \tau \tau\right)$ as a black line and constraints due to the
measured $\eta_b(1S)$ mass by Babar as a red line;
from\cite{Domingo:2008rr}.} 
\end{figure}

On the other hand, the average value for
the hyperfine splitting $E_{hfs}(1S) = m_{\Upsilon(1S)} - 
m_{\eta_b(1S)}$ found by BABAR~\cite{:2009pz} is somewhat large:
\begin{equation}\label{ehfsexp}
E_{hfs}^{exp}(1S) = 69.9 \pm 3.1\ \mathrm{MeV}
\end{equation}
This result can be compared to recent  predictions from  perturbative
QCD: $E_{hfs}(1S) = 44 \pm 11\ \mathrm{MeV}$  \cite{Recksiegel:2003fm}
and $E_{hfs}(1S) = 39 \pm 14\ \mathrm{MeV}$ \cite{Kniehl:2003ap}.

Whereas an explanation of the discrepancy between (\ref{ehfsexp}) and
perturbative QCD is not excluded at present, the difference might be
ascribed to the mixing of the  $\eta_b(1S)$ state with a light
pseudoscalar Higgs $A_1$, if $M_{A_1}$ is somewhat above 9.4~GeV. (For
$M_{A_1} > 9$~GeV, $X_d$ is bounded from above just by the red line in
Fig.~2.) The assumption that the $\eta_b(1S)$-$A_1$ mixing explains the
discrepancy between (\ref{ehfsexp}) and perturbative QCD, allows to
determine $X_d$ and hence all eigenvalues and mixing angles in the
$\eta_b(nS)$-$A_1$-system as function
of~$M_{A_1}$~\cite{Domingo:2009tb}.  Then, the masses of the states
interpreted as $\eta_b(2S)$ and $\eta_b(3S)$ can also be modified.
Furthermore, all $\eta_b(nS)$ states can acquire non-negligible
branching ratios into $\tau^+\,\tau^-$ due to their mixing with $A_1$.

In Fig.~3 (from \cite{Domingo:2009tb}), the masses of all 4 physical
states (denoted by $\eta_i$, $i=1\dots 4$) as functions of $M_{A_1}$ 
are shown together with the error bands; by construction, $m_{\eta_1}
\equiv 9.39$~GeV and for clarity the assumed values for
$m_{\eta_b^0(nS)}$ (before mixing) are indicated as horizontal dashed
lines. For
$M_{A_1}$ not far above 9.4~GeV the effects of the mixing on the states
$\eta_b^0(2S)$ and $\eta_b^0(3S)$ are negligible, but for larger
$M_{A_1}$ the spectrum can differ considerably from the expectations
from QCD-based quark models.
\begin{figure}
\begin{center}
\resizebox{0.35\textwidth}{!}{%
  \includegraphics{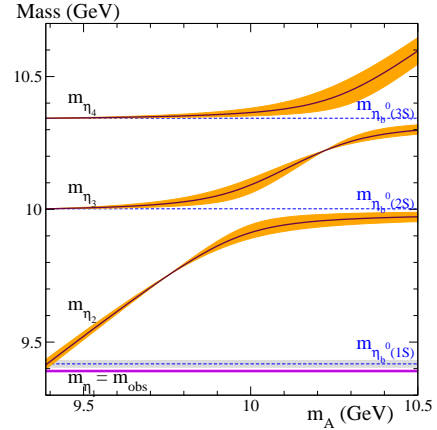}}
%\addvspace{-6mm}
\caption{The masses of all eigenstates of the $\eta_b(nS)$-$A_1$-system
as function of $M_{A_1}$, once the $\eta_b(1S)$-$A_1$-mixing is required
to explain the discrepancy between (\ref{ehfsexp}) and perturbative
QCD.}
\end{center}
\label{fig:heavy_masses}
\end{figure}

Future experiments at $B$ factories could verify this scenario in
different ways: through the spectrum of the $\eta_b(nS)$-$A_1$-system
beyond the $\eta_b(1S)$-state seen by BABAR, and/or through a violation
of lepton universality in inclusive $\Upsilon$ decays $\Upsilon \to
l^+\,l^- +X$: from $\Upsilon \to \gamma + A_1 \to \gamma +
\tau^+\,\tau^-$ one would expect an excess in the $\tau^+\,\tau^-$ final
state~\cite{SanchisLozano:2002pm,Domingo:2008rr}.

\subsection{Higgs Searches at the LHC with dominant $H\to A_1 A_1$
decays}

If a SM-like Higgs boson $H$ decays dominantly into a pair of light
pseudoscalars with $M_{A_1} < 10.5$~GeV (such that $A_1$ decays
dominantly into $\tau^+\,\tau^-$), the detection of $H$ will be quite
difficult at the LHC: the final state from $H\to A_1 A_1 \to 4\tau$ will
contain four $\tau$ neutrinos, which escape undetected and make it
difficult to observe peaks in invariant masses corresponding to $A_1$ or
$H$; two $\tau$ leptons at a time (originating from one $A_1$ boson)
will be nearly collinear, and the average $p_T$ of the particles in the
final state is quite low. Important backgrounds originate from $\Psi$
production and heavy flavour jets.

Up to now, the following proposals for $H$ searches at the LHC in this
scenario have been made:

(i) In \cite{Forshaw:2007ra} it has been proposed to consider
diffractive Higgs production ($pp \to pp+H$) in order to be sensitive to
$H \to 4\tau$, which requires to install additional forward detectors.
Using a track-based analysis in which all events with more than 6 tracks
in the central region are discarded, a viable signal seems possible
after accumulating 300~fb$^{-1}$ of integrated luminosity.

(ii) Proposals for signals and cuts appropriate for the $A_1\,A_1 \to
4\tau \to 2\,\mu + 2\,$jets final state have been made in
\cite{Belyaev:2008gj}; with 100~fb$^{-1}$ of integrated luminosity, the
expected rates after cuts are $\sim 8\cdot 10^3$ from  $H$~production
via vector boson fusion, and $\sim 10^3$ from $H$~production via Higgs
Strahlung ($W^{\pm\ *} \to H+W^\pm$) where one can trigger on a lepton
from $W^\pm$ decays.

(iii) In \cite{Lisanti:2009uy}, the subdominant $H \to A_1\,A_1 \to
2\tau\,2\mu$ final state (with $2\mu$ from direct $A_1$ decays)  was
discussed: in spite of the small branching fraction it was argued that,
for $M_H \sim 102$~GeV and with $H$ being produced via gluon-gluon
fusion, the Tevatron can see a signal over the background for an
integrated luminosity ${\cal L}\sim 10$~fb$^{-1}$, and the LHC already
for ${\cal L} \sim 1$~fb$^{-1}$.

Further details of current ATLAS and CMS studies of bench mark scenarios
including the $H \to A\,A \to 4\tau$ final state can be found
in~\cite{Djouadi:2008uw}.

\section{The constrained NMSSM}

As in the MSSM, one can assume universal soft SUSY breaking masses and
trilinear couplings at the GUT or Planck scale, which is motivated by a
gravitational (flavour blind) origin for these terms in minimal
supergravity. Then, all soft SUSY breaking terms (including those
involving the singlet in the NMSSM) are specified by universal gaugino
masses $M_{1/2}$, universal scalar masses $m_0$ and universal trilinear
couplings $A_0$ at a large scale. The constrained NMSSM (cNMSSM) has the
same number of free parameters as the constrained MSSM (cMSSM): the
parameters $\mu$, $B$ of the cMSSM are replaced by the Yukawa couplings
$\lambda$, $\kappa$ of the cNMSSM.

The allowed ranges of the parameters $M_{1/2}$, $m_0$ and $A_0$ in the
cMSSM have been widely discussed in the literature; it turns out that
these ranges are very different in the
cNMSSM~\cite{Djouadi:2008yj,Djouadi:2008uj}: in order to obtain a
non-vanishing vev
$\left<S\right> \neq 0$ in the NMSSM, the soft SUSY breaking mass
$m_S^2$ must not be large and positive. Since $m_S^2$ is hardly
renormalized between the GUT and the weak scale, the same condition
applies to $m_0^2$ in the cNMSSM. In the cMSSM, small values of $m_0$
(compared to $M_{1/2}$) lead to a charged stau ($\widetilde{\tau}$) LSP,
which is ruled out. In the cNMSSM, an additional singlino-like
neutralino $\chi_1^0$ can have a mass below the $\widetilde{\tau}$~mass
and be the true LSP. In order to give the correct (not too large) dark
matter relic density, $M_{\chi_1^0}$ must be just a few GeV below
$M_{\widetilde{\tau}}$, such that the $\chi_1^0$ relic density can be
reduced via co-annihilation with the $\widetilde{\tau}$ NLSP. The latter
condition fixes $A_0$ in terms of $M_{1/2}$. Finally LEP constraints on
the Higgs sector lead to an upper bound on $\lambda$ of $\lambda \lsim
0.02$~\cite{Djouadi:2008yj,Djouadi:2008uj}.

\begin{figure}[h!]
\includegraphics[scale=0.27]{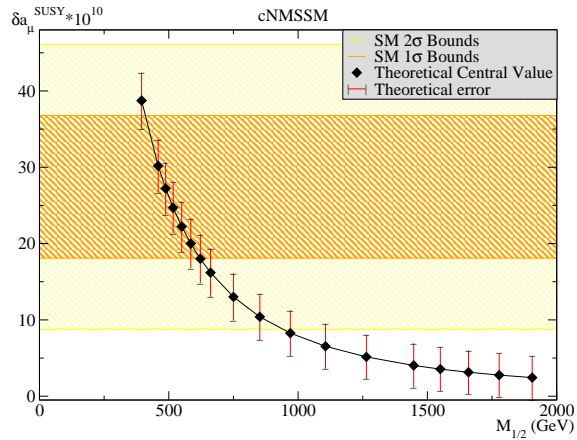} \caption{$\delta
a_\mu^\mathrm{SUSY}$ as a function of $M_{1/2}$ in the cNMSSM;
from\cite{Domingo:2008bb}.} 
\label{CNMSSMfig}
\end{figure}

Hence, the full Higgs and sparticle spectrum depends essentially on
$M_{1/2}$ only. A preferred range for $M_{1/2}$ can be obtained by
computing the muon anomalous magnetic moment $a_\mu$ as a function of
$M_{1/2}$, and requiring that the supersymmetric contributions
explain the deviation $\delta a_\mu \sim 3\times 10^{-9}$ between the
result of the E821 experiment at BNL~\cite{Bennett:2006fi} and the
SM~\cite{Domingo:2008bb}. The result for $\delta a_\mu^\mathrm{SUSY}$ is
shown in Fig.~\ref{CNMSSMfig}, according to which values for $M_{1/2}
\lsim 1$~TeV are preferred, with $M_{1/2} \approx 500$~GeV within
$1\,\sigma$.

\begin{figure}[hb!]
\includegraphics[scale=0.35,angle=-90]{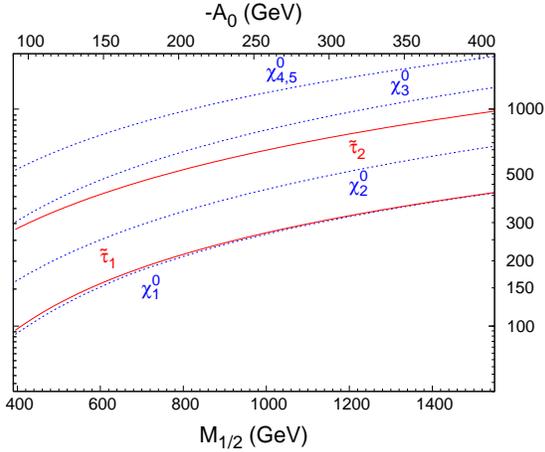} 
\caption{The stau and neutralino spectrum as a function of $M_{1/2}$ in
the cNMSSM; from~\cite{Djouadi:2008yj}.} 
\label{phenoX0fig}
\end{figure}

Choosing $m_0 = 0$ for simplicity, the stau and neutralino spectrum as a
function of $M_{1/2}$ in the cNMSSM is shown in Fig.~\ref{phenoX0fig}.
One finds that, for $M_{1/2} \lsim 400$~GeV, the lightest stau mass
would fall below $\sim 100$~GeV and violate LEP constraints; hence one
must require $M_{1/2} \gsim 400$~GeV. (As
discussed above, the $\chi_1^0$ mass is just below the lightest stau
mass in order to give the correct dark matter relic density.)

\begin{figure}
\includegraphics[scale=0.35,angle=-90]{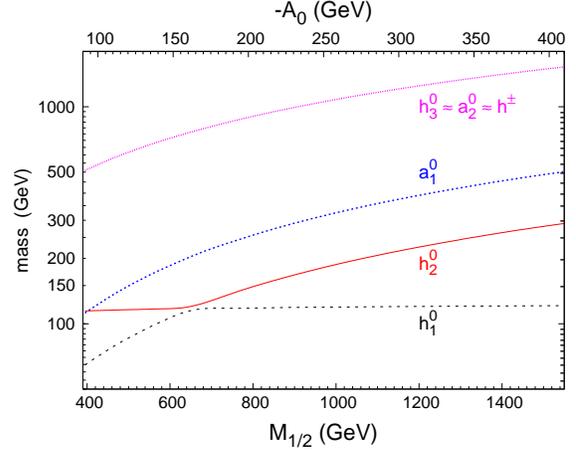} 
\caption{The Higgs spectrum as a function of $M_{1/2}$ in
the cNMSSM; from~\cite{Djouadi:2008yj}.} 
\label{phenoHfig}
\end{figure}

The Higgs spectrum as a function of $M_{1/2}$ in the cNMSSM is shown in
Fig.~\ref{phenoHfig}. The heavy states $H_3$, $A_2$ and $H^\pm$ form a
practically degenerate $SU(2)$ doublet, and the lighter CP-odd scalar
$A_1$ is too heavy to be produced in $H_1$ decays in the cNMSSM. The
nature of the lightest CP-even Higgs scalar depends on $M_{1/2}$: for
$M_{1/2} \lsim 640$~GeV, $H_1$ is dominantly singlet-like which allows
to satisfy LEP constraints even for a mass well below 114~GeV. The
next-to-lightest state $H_2$, for $M_{1/2} \lsim 640$~GeV, is SM-like
with a mass just above 115~GeV. For $M_{1/2} \gsim 640$~GeV, the
situation is reversed: here $H_2$ is dominantly singlet-like, and the
lightest CP-even Higgs scalar $H_1$ is SM-like with a mass up to 120~GeV
for $M_{1/2} \to 1.5$~TeV.

An interesting scenario is possible for $M_{1/2} \sim 570$~GeV, where
$M_{H_1} \sim 100$~GeV and the reduced coupling $\xi_{H_1 ZZ}$ of the
singlet-like $H_1$ (for not too small $\lambda \sim 0.005$) is $\xi_{H_1
ZZ}\sim 0.3$~\cite{Djouadi:2008uj}: in this case a production of $H_1$
at LEP could explain the excess of events at LEP as discussed before.
Simultaneously, $H_2$ with its mass just above 114~GeV could generate
an even slighter excess of events in this mass range, which has equally
been observed~\cite{Schael:2006cr}.

The consequences of the nearly pure singlino-like $\chi_1^0$ LSP in the
cNMSSM on sparticle decay chains are important: first, each sparticle
decay cascade will end up in the next-to-lightest $R$-odd particle, the
charged $\widetilde{\tau}$ in the present case. Only subsequently the
$\widetilde{\tau}$ will decay as $\widetilde{\tau} \to \tau + \chi_1^0$,
leading to at least one $\tau$ lepton per sparticle. Notably, if the
$\widetilde{\tau} - \chi_1^0$ mass difference is small and/or $\lambda$
is very small (in which case the couplings to $\chi_1^0$ become tiny),
the $\widetilde{\tau}$ width can become so small that the
$\widetilde{\tau}$ decay length becomes ${\cal
O}$(mm-cm)~\cite{Djouadi:2008yj,Djouadi:2008uj}. Such displaced vertices
can also appear in scenarios with gauge mediated supersymmetry breaking
where, however, the decay lengths are rather of ${\cal O}$(m). Hence,
dedicated simulations of the sparticle signatures at colliders in the
cNMSSM will be required.

\section{Summary and Conclusions}

Assuming that the SUSY breaking scale $M_\mathrm{SUSY}$ generates the
weak scale $\sim M_Z$ (i.e. that no dimensionful terms as $\mu$ are
present in the superpotential), the NMSSM is the most natural
supersymmetric extension of the Standard Model.

Its simplest version, the cNMSSM (with soft SUSY breaking terms from
mSUGRA), could explain the $2.3\ \sigma$ excess of events in Higgs
searches at LEP, in contrast to the cMSSM. Due to $m_0 \ll M_{1/2}$, its
sparticle spectrum would be very different from the cMSSM: the LSP
$\chi_1^0$ is singlino-like, but every sparticle decay chain contains a
$\widetilde{\tau}$ decaying, in turn, into $\tau + \chi_1^0$. This last
decay can lead to macroscopically displaced vertices.

In the general NMSSM, a light CP-odd scalar $A_1$ with $M_{A_1} <
10.5$~GeV could (i) explain the $2.3\ \sigma$ excess of events at
LEP; (ii) alleviate the ``little finetuning problem'' of supersymmetric
extensions of the SM, and (iii) explain a low $\eta_b$ mass as measured
by BABAR. But: this scenario would constitute a real challenge for Higgs
searches both at the Tevatron and at the LHC!


\begin{thebibliography}{99}
  
\bibitem{Kim:1983dt}
  J.~E.~Kim and H.~P.~Nilles,
  %``The Mu Problem And The Strong CP Problem,''
  Phys.\ Lett.\  B {\bf 138} (1984) 150.
 
\bibitem{LEPSUSYWG}
LEPSUSYWG, ALEPH, DELPHI, L3 and OPAL experiments,
note  LEPSUSYWG/01-03.1 
%{\sf (http://lepsusy.web.cern.ch/lepsusy/Welcome.html)}. 
  
\bibitem{Schael:2006cr}
  S.~Schael {\it et al.}  [ALEPH Collaboration and DELPHI Collaboration and
                  L3 Collaboration and ],
  %``Search for neutral MSSM Higgs bosons at LEP,''
  Eur.\ Phys.\ J.\  C {\bf 47} (2006) 547
  [arXiv:hep-ex/0602042].

\bibitem{Dermisek:2005gg}
  R.~Dermisek and J.~F.~Gunion,
  %``Consistency of LEP event excesses with an h --> a a decay scenario and
  %low-fine-tuning NMSSM models,''
  Phys.\ Rev.\  D {\bf 73} (2006) 111701
  [arXiv:hep-ph/0510322].

\bibitem{Dermisek:2006wr}
  R.~Dermisek and J.~F.~Gunion,
  %``The NMSSM close to the R-symmetry limit and naturalness in h --> aa  decays
  %for m(a) < 2m(b),''
  Phys.\ Rev.\  D {\bf 75} (2007) 075019 [arXiv:hep-ph/0611142].
  
\bibitem{:2008hs}
  W.~Love {\it et al.}  [CLEO {Collaboration}],\newline
  %``Search for Very Light CP-Odd Higgs Boson in Radiative Decays of
  %Upsilon(S-1),''
  Phys.\ Rev.\ Lett.\  {\bf 101} (2008) 151802
  [arXiv: 0807.1427 [hep-ex]].

\bibitem{Aubert:2008vj}
   B.~Aubert  {\it et al.}  [BABAR Collaboration], Phys.\ Rev.\ Lett.\ 
  {\bf 101} (2008) 071801 [arXiv: 0807.1086 [hep-ex]].

\bibitem{:2009pz}
  B.~Aubert {\it et al.}  [BABAR Collaboration],\newline
  %``Evi\-dence for the $\eta_b(1S)$ Meson in Radiative $\Upsilon(2S)$ 
  %Decay,''
  arXiv:0903.1124 [hep-ex].

\bibitem{Drees:1989du}
  M.~Drees and K.-i.~Hikasa,
  %``HEAVY QUARK THRESHOLDS IN HIGGS PHYSICS,''
  Phys.\ Rev.\  D {\bf 41} (1990) 1547.
  
\bibitem{Fullana:2007uq}
  E.~Fullana and M.~A.~Sanchis-Lozano,\nl
  %``Hunting a light CP-odd non-standard Higgs boson through its tauonic   decay
  %at a (Super) B factory,''
  Phys.\ Lett.\  B {\bf 653} (2007) 67
  [arXiv:hep-ph/0702190].
  
\bibitem{Domingo:2008rr}
  F.~Domingo, U.~Ellwanger, E.~Fullana,\nl
   C.~Hugonie and M.~A.~Sanchis-Lozano,
  %``Radiative Upsilon decays and a light pseudoscalar Higgs in the NMSSM,''
  JHEP {\bf 0901} (2009) 061
  [arXiv:0810.4736 [hep-ph]].
  
\bibitem{Domingo:2009tb}
  F.~Domingo, U.~Ellwanger and M.~A.~Sanchis-Lozano,
  %``Bottomoniom spectroscopy with mixing of eta_b states and a light CP-odd
  %Higgs,''
  arXiv:0907.0348 [hep-ph], to appear in Phys.\ Rev.\ Lett.
  
\bibitem{Recksiegel:2003fm}
  S.~Recksiegel and Y.~Sumino,
  %``Fine and hyperfine splittings of charmonium and bottomonium: 
  %An improved perturbative QCD approach,''
  Phys.\ Lett.\  B {\bf 578} (2004) 369
  [arXiv:hep-ph/0305178].
  
\bibitem{Kniehl:2003ap}
  B.~A.~Kniehl, A.~A.~Penin, A.~Pineda,\nl
   V.~A.~Smirnov and M.~Steinhauser,
  %``M(eta/b) and alpha(s) from nonrelativistic renormalization group,''
  Phys.\ Rev.\ Lett.\  {\bf 92} (2004) 242001
  [arXiv:hep-ph/ 0312086]. 

\bibitem{SanchisLozano:2002pm}
  M.~A.~Sanchis-Lozano,
  %``Searching for new physics in leptonic decays of bottomonium,''
  Mod.\ Phys.\ Lett.\  A {\bf 17} (2002) 2265
  [arXiv:hep-ph/0206156]  
  
\bibitem{Forshaw:2007ra}
  J.~R.~Forshaw,
  % {\it et al.}, 
  J.~F.~Gunion, L.~Hodgkinson, A.~Papaefstathiou and A.~D.~Pilkington,
  %``Reinstating the 'no-lose' theorem for NMSSM Higgs discovery at the LHC,''
  JHEP {\bf 0804} (2008) 090\newline
  [arXiv:0712.3510 [hep-ph]].

\bibitem{Belyaev:2008gj}
  A.~Belyaev
  {\it et al.},
 %S.~Hesselbach, S.~Lehti, S.~Moretti,  A.~Nikitenko and
 %C.~H.~Shepherd-Themistocleous,
  %``The Scope of the 4 tau Channel in Higgs-strahlung and Vector Boson Fusion
  %for the NMSSM No-Lose Theorem at the LHC,''
  %\newline 
  arXiv:0805.3505 [hep-ph].
  
\bibitem{Lisanti:2009uy}
  M.~Lisanti and J.~G.~Wacker,
  %``Discovering the Higgs with Low Mass Muon Pairs,''
  Phys.\ Rev.\  D {\bf 79} (2009) 115006
  [arXiv:0903.1377 [hep-ph]].
 
\bibitem{Djouadi:2008uw}
  A.~Djouadi {\it et al.},
  %``Benchmark scenarios for the NMSSM,''
  JHEP {\bf 0807} (2008) 002
  [arXiv:0801.4321 [hep-ph]].

\bibitem{Djouadi:2008yj}
  A.~Djouadi, U.~Ellwanger and A.~M.~Teixeira,
  %``The constrained next-to-minimal supersymmetric standard model,''
  Phys.\ Rev.\ Lett.\  {\bf 101} (2008) 101802
  [arXiv:0803.0253 [hep-ph]].
  
\bibitem{Djouadi:2008uj}
  A.~Djouadi, U.~Ellwanger and A.~M.~Teixeira,
  %``Phenomenology of the constrained NMSSM,''
  JHEP {\bf 0904} (2009) 031
  [arXiv:0811.2699 [hep-ph]].
 
\bibitem{Bennett:2006fi}
  G.~W.~Bennett {\it et al.}  [Muon G-2 Collaboration],
  %``Final report of the muon E821 anomalous magnetic moment measurement at
  %BNL,''
  Phys.\ Rev.\  D {\bf 73} (2006) 072003
  [arXiv:hep-ex/0602035].
   
\bibitem{Domingo:2008bb}
  F.~Domingo and U.~Ellwanger,
  %``Constraints from the Muon g-2 on the Parameter Space of the NMSSM,''
  JHEP {\bf 0807} (2008) 079
  [arXiv:0806.0733 [hep-ph]].

\end{thebibliography}
\end{document}